\documentstyle[prd,epsfig,aps,floats]{revtex} 

\draft

\begin{document}
\twocolumn[\hsize\textwidth\columnwidth\hsize\csname
@twocolumnfalse\endcsname
\title{%
\hbox to\hsize{\normalsize\rm YARU-HE-98/02
\hfil MPI-PTh/98-25}
\vskip 40pt
Axion Emission by Magnetic-Field Induced
Conversion of Longitudinal Plasmons}
\author{N.V.~Mikheev}
\address{Department of Theoretical Physics, Yaroslavl State 
University, Sovietskaya 14, Yaroslavl 150000, Russia}
\author{G.~Raffelt}
\address{Max-Planck-Institut f\"ur Physik 
(Werner-Heisenberg-Institut),
F\"ohringer Ring 6, 80805 M\"unchen, Germany}
\author{L.A.~Vassilevskaya}
\address{Moscow State (Lomonosov) University, V-952, Moscow 117234,
Russia}

\date{\today}

\maketitle

\begin{abstract}
Magnetic fields mix axions with photons, allowing for the cyclotron
process $e^-\to e^- a$ by virtue of an intermediate plasmon even if
axions do not couple to electrons at tree level.  The axion and
longitudinal-plasmon dispersion relations always cross for a certain
wave-number, leading to a resonant enhancement of this process. Even
then, however, it cannot quite compete with the usual nucleon
processes in a supernova core.  The well-known axion window
$10^{-5}~{\rm eV}\alt m_a\alt 10^{-2}~{\rm eV}$ remains open and
axions may still constitute the cosmic dark matter.
\end{abstract}
\vskip2.0pc]


\section{Introduction}

Twenty years after their invention, axions still remain the most
elegant solution of the strong CP problem~\cite{P1,WW}, but alas, they
also remain undiscovered.  In terms of their mass, which almost
uniquely characterizes a given axion model, the well-known ``window of
opportunity'' $10^{-5}~{\rm eV}\alt m_a\alt 10^{-2}~{\rm eV}$ has been
left open by laboratory searches and by astrophysical and cosmological
arguments~\cite{Tur,Raf1,Raf2}. Somewhere in this range axions could
well be the main component or a significant fraction of the cosmic
dark matter. The ongoing search experiments for galactic dark-matter
axions~\cite{Kyoto,Livermore} offer our best chance to discover the
existence of these ``invisible'' particles.

Meanwhile it remains of great interest to look for other astrophysical
sites or laboratory experiments where the existence of axions could
manifest itself. In the astrophysical context, relatively little
attention has been paid to axionic processes in strong magnetic fields
even though field strengths far in excess of the electron's Schwinger
value of $B_e=m^2_e/e\simeq4.41\times10^{13}~{\rm Gauss}$ may well
exist in the interior of supernova cores or pulsars.  Possible
mechanisms to generate fields as strong as $10^{15}$--$10^{17}~{\rm
Gauss}$ in stars~\cite{magnetar,tor} or the early
universe~\cite{Lem,Taj} are now widely discussed.  Magnetic fields
enable processes which are otherwise forbidden. An example is the
cyclotron emission of axions $e^-\to e^- a$ in meutron stars or
magnetic white dwarfs~\cite{Kachelriess}, even though the field
strengths required to obtain an observable effect seem unrealistically
large.

We here study a related process which is induced by a strong magnetic
field, the axion cyclotron emission $e^-\to e^-a$ via a plasmon
intermediate state. Therefore, the axion coupling is to the
electromagnetic field so that our process does not depend on a direct
axion-electron coupling which exists only in a restricted class of
models~\cite{DFSZ}. The main motivation why our process could be
expected to be non-negligible is that it is resonant at a particular
energy of the emitted axion, provided that the intermediate plasmon is
longitudinal, because the axion and the longitudinal plasmon
dispersion relations always cross for a certain wave-number (Fig.~1).

\begin{figure}[b]
\center\leavevmode
\epsfxsize=6.2cm
\epsfbox{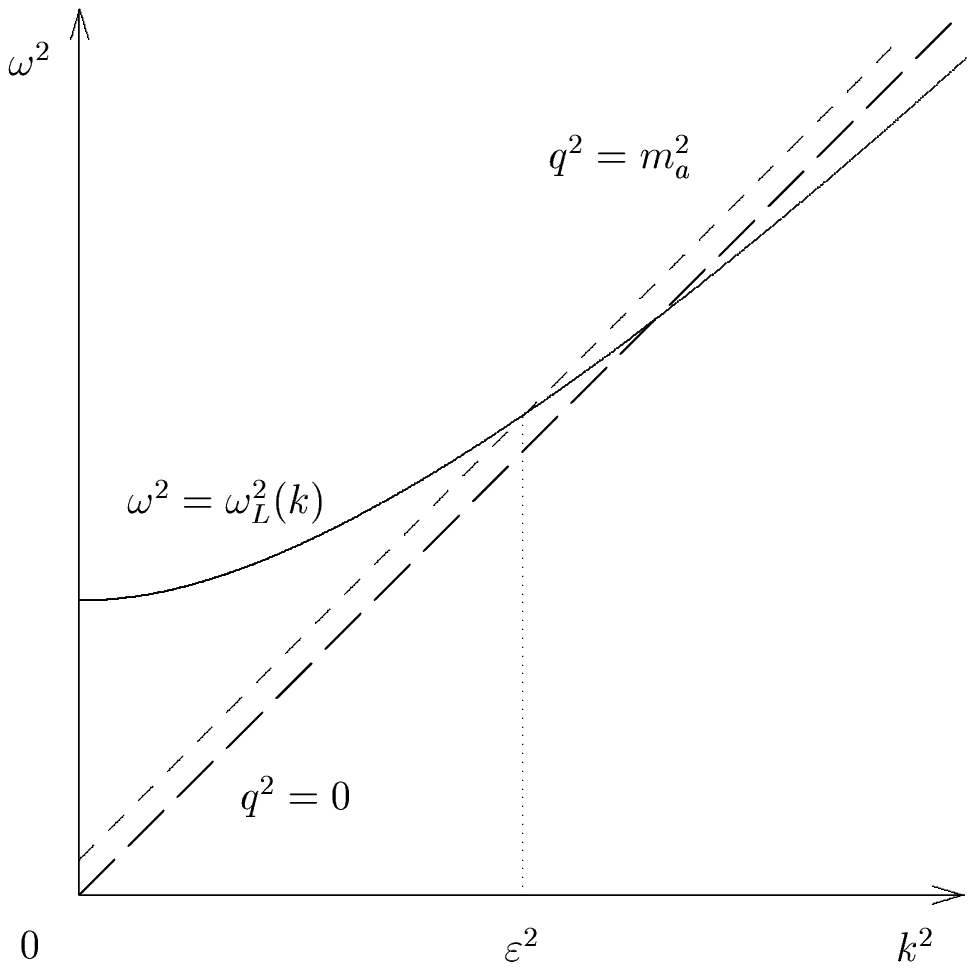}
\medskip
\caption{Dispersion relation $\omega^2=\omega^2_L(k)$ for
longitudinal plasmons (solid line), axions $\omega^2 = k^2+m^2_a$
(short dashes), and vacuum photons $\omega=k$ (long dashes).}
\end{figure}

Our final result will be that even for the largest magnetic fields
that may plausibly exist in supernova cores this new emission process
is less important than the usual nucleon bremsstrahlung
rate. Therefore, the axion ``window of opportunity'' remains
unaffected by our process.  While from a practical astrophysical
perspective our results turn out to be purely academic we still find
it worthwhile to communicate this calculation for its 
conceptual~value.

In Sec.~II we calculate the emission rate when our effect is pictured
as a cyclotron process $e^-\to e^- a$. In Sec.~III we calculate it
again, picturing it as a plasmon-axion oscillation phenomenon. Because
we are on resonance, both calculations yield the same result.  In
Sec.~IV we summarize our findings.


\section{Cyclotron Process}

\subsection{Axion-Photon Coupling}

The cyclotron process $e^-\to e^- a$ via an intermediate plasmon is
shown in Fig.~2, where double lines correspond to the electron wave
functions in the magnetic field. The effective Lagrangian describing
the axion-photon coupling in the presence of an external field can be
presented as
\begin{eqnarray}
{\cal L}_{a \gamma}=\bar g_{a\gamma}\,\partial_{\mu}A_\nu\,
\tilde F_{\mu\nu}\,a,
\label{eq:Lag} 
\end{eqnarray}
where $A_\mu$ is the four potential of the quantized electromagnetic
field, $\tilde F$ is the dual of the tensor which describes the
external field, and $a$ the axion field.

\begin{figure}[b]
\center\leavevmode
\epsfxsize=4.0cm
\epsfbox{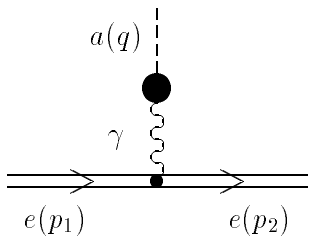}
\bigskip
\caption{Cyclotron emission process.}
\end{figure}

The effective $a\gamma$ coupling constant with the dimension
(energy)$^{-1}$ is
\begin{eqnarray}
\bar g_{a\gamma} =  g_{a \gamma} + \frac{\alpha}{\pi} \, 
\sum_{{\rm light~}f} \frac{Q_f^2 g_{af}}{m_f}\,.
\label{eq:Gag1}
\end{eqnarray}
Here $g_{a\gamma}$ is the usual coupling constant in
vacuum~\cite{Raf1} $g_{a\gamma}=\alpha\xi/2\pi f_a$ where $\xi$ is a
model-dependent parameter and $f_a$ the Peccei-Quinn scale.  Further,
$g_{af} = C_f m_f/f_a$ is a dimensionless Yukawa coupling constant of
axions to both quarks and leptons at tree level with $C_f$ a
model-dependent factor, $m_f$ the fermion mass, and $Q_f$ its relative
electric charge.  The second term in Eq.~(\ref{eq:Gag1}) is the
field-induced part of the coupling which derives from the diagram of
Fig.~3.  It has contributions from light fermions $f$ for which
$\chi_f\gg 1$.  The dynamic parameter is defined as
\begin{equation}
\chi_f^2 = \frac{e_f^2 (qFFq)}{m_f^6}
\label{eq:Chi}
\end{equation}
with $e_f=Q_f e$ the fermion electric charge and $q$ is the axion 
four-momentum.

\begin{figure}[t]
\center\leavevmode
\epsfxsize=4.0cm
\epsfbox{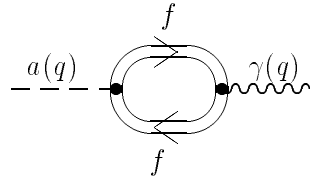}
\bigskip
\caption{Magnetic-field induced modification of the axion-photon
coupling.}
\end{figure}

The effective axion-photon coupling constant in the presence of an
external field is then written as
\begin{equation}
\bar g_{a\gamma}=\frac{\alpha}{2 \pi f_a}\,
\biggl(\xi+2\sum_{{\rm light~} f} Q_f^2 C_f\biggr).
\label{eq:Gag2}
\end{equation}
Note that for ultrarelativistic electrons ($E \gg m_e$) the case
$E^2\gg eB$, when a large number of the Landau levels are excited, is
well described by the crossed-field limit $(e^2 p F F p)^2 \gg (e^2 F
F)^3$.

\subsection{Matrix Element}

The matrix element of the $e^- \to e^- a$ decay, corresponding to the
diagram of Fig.~2, can be written as
\begin{equation}\label{eq:S1}
S=\frac{\bar g_{a\gamma}}{\sqrt{2 E_a V}} \, h J
\end{equation}
in terms of the currents
\begin{eqnarray}
J_\alpha&=&\int d^4 x\,
{\bar\psi(p_2,x)}\,\gamma_{\alpha}\,\psi(p_1,x)\,e^{iqx},
\nonumber\\
h_{\alpha}&=&
- i e q_\mu \tilde F_{\mu\nu} G^{L}_{\nu\alpha}(q).
\end{eqnarray}
Here, $e>0$ is the elementary charge, $q = (E_a, {\bf q})$ is the
four momentum of the emitted axion, and $p_1=(E_1,{\bf p}_1)$ and
$p_2=(E_2,{\bf p}_2)$ are the quasi-momenta in the crossed field of
the initial- and final-state electrons with $p_1^2=p_2^2=m_e^2$; in
the zero-field limit $p_{1,2}$ become the free electron four-momenta.

\eject

Further, $\psi(p,x)$ is the solution of the Dirac equation in a
constant crossed field $F_{\mu\nu}=k_\mu a_\nu-k_\nu a_\mu$ where
$A_\mu = a_\mu \varphi$ is the four-potential with $\varphi=kx$ and
$k^2=ak=0$. We have explicitly
\begin{equation}\label{eq:Psi}
\psi(p,x)=\biggl(1-\frac{e\,\hat k \hat a}{2\,kp}\,\varphi\biggr) 
\,\frac{U (p)}{\sqrt{2 E V} }\,e^{ i I(p,x)},
\end{equation}
where $\hat k\equiv\gamma_\mu k_\mu$ and 
\begin{equation}
I (p,x) = -p x + \frac{e\, a p}{2\, kp}\,\varphi^2 +
\frac{e^2 a^2}{6\,k p}\,\varphi^3.
\end{equation}
The bispinor $U(p)$, which is normalized by the condition $\bar U U =
2 m_e$, satisfies the Dirac equation for the free electron $(\hat
p-m_e)U(p)=0$.

Finally, $G^{L}_{\alpha\beta}$ is the longitudinal plasmon propagator.
Because the magnetic field is taken to be weak on the scale of the
medium temperature $T$ and electron Fermi momentum $p_F$ 
($T^2,~p_F^2\gg eB$) one can use the zero-field propagator
\begin{equation}\label{eq:G}
G^{L}_{\alpha\beta}=i\,\frac{\ell_\alpha\ell_\beta}{q^2-\Pi^L}.
\end{equation}
Here, $\ell_\alpha$ and $\Pi^{L}$ are the eigenvector and eigenvalue
of the polarization operator corresponding to the longitudinal
plasmon, respectively, with
\begin{equation}
\ell_\alpha=\sqrt{\frac{q^2}{(uq)^2 - q^2}}\,
\left(u_\alpha - \frac{uq}{q^2}\,q_\alpha \right)
\nonumber
\end{equation}
and $u_{\alpha}$ the four-velocity of the medium.

The general expression for the matrix element~Eq.~(\ref{eq:S1}) is
rather cumbersome. It is substantially simplified in the
ultrarelativistic limit where $E_{1,2} \gg m_e$. After integration
over the space-time coordinates $x$ it is
\begin{eqnarray}\label{eq:S2}   
S&\simeq&\frac{(2 \pi)^4 \delta^2({\bf Q}_{\perp})\, 
\delta(\chi_1 - \chi_2 - \chi_q)}   
{\sqrt{2 E_a V\,2 E_1 V\, 2 E_2 V}}\nonumber\\
&\times&\frac{\bar g_{a\gamma}}{\pi m_e^2 r}\,\Phi(\eta)\, 
\bar U(p_2)\,\hat h\,U(p_1),
\end{eqnarray}
where $Q=p_1-p_2-q$ is the four-momentum transfer to the external
field and $Q_\perp$ its perpendicular component defined by the
condition ${\bf Q}_{\perp}\cdot{\bf k}=0$.
Further, $r=(\chi_a/2 \chi_1 \chi_2)^{1/3}$, 
$\chi_i^2 =e^2 (p_i F F p_i)/m_e^6$ for $i=1,2,a$ 
with $p_a=q$ the axion momentum, and 
$\eta=r^2(1 + \tau^2)$ with
$\tau=-e(p_1 \tilde F q)/(m_e^4 \chi_a)$. 
Finally,
\begin{equation}\label{eq:Ai}
\Phi(\eta)= \int_0^\infty dy\, \cos \left ( \eta y +
\frac{y^3}{3}\right)
\end{equation}
is the Airy function.

\subsection{Resonant Transition Rate}

After integrating over the final-state electron momentum and the axion
solid angle, the differential probability of $e^-\to e^-a$ is
\begin{equation}\label{eq:W1}
\frac{dW}{dE_a}\simeq\frac{\bar g_{a\gamma}^2}{12 \pi}\,(e B)^2\,
\frac{E_a^2\,\cos^2\theta}{({\cal E}^2-E_a^2)^2+\gamma^2{\cal E}^4}\, 
\left(1-\frac{E_a}{E_1}\right),
\end{equation}
where $\theta$ is the angle between the external magnetic field ${\bf
B}$ and the momentum ${\bf p}_1$ of the initial electron.  ${\cal E}$
is the energy where the axion and longitudinal plasmon dispersion
curves cross. In an ultrarelativistic degenerate plasma it is
\begin{equation}\label{eq:Dis2}
{\cal E}^2 \simeq \frac{4 \alpha}{\pi}\,\mu^2\,
\left(\ln\frac{2\mu}{m_e}-1\right),
\end{equation}
where $\mu$ is the electron chemical potential.

The dimensionless resonance width $\gamma$ of the $e^-\to
e^-a$ process in Eq.~(\ref{eq:W1}) is 
\begin{equation}\label{eq:gamma}
\gamma=\frac{{\cal E} \Gamma_L({\cal E})}{q^2 Z_L},
\end{equation}
where at the resonance point, of course, $q^2=m_a^2$. The plasmon
wave-function renormalization factor is
\begin{equation}\label{eq:Gamma1}
Z_L^{-1} = 1 - \frac{\partial \,\Pi^{(L)}}{\partial\,q_0^2 }.
\end{equation} 
In an ultrarelativistic degenerate plasma it is
\begin{eqnarray}
Z_Lq^2\simeq2{\cal E}^2\,\frac{m^2_e[\ln(2\mu/m_e) - 1]}{\mu^2}
\label{eq:ZL}
\end{eqnarray} 
with the electron chemical potential $\mu$.

Further, $\Gamma_L ({\cal E})$ is the total width of the longitudinal
plasmon due to the processes $\gamma_L\to\nu\bar\nu$, $\gamma_L\to
e^+e^-$, $\gamma_L e^- \to e^-$, and so forth in the presence of the
magnetic field.  It turns out that the main contribution comes from
the inverse cyclotron process $\gamma_L e^- \to e^-$.  A direct
calculation of the absorption rate due to this process yields
\begin{eqnarray} 
\Gamma_{\rm ab}({\cal E})\simeq\frac{2 \alpha}{3}\,q^2 Z_L\, 
\frac{\mu^2}{{\cal E}^3}\,
\frac{e^{{\cal E}/T}}{e^{{\cal E}/T} - 1}.
\label{eq:G1} 
\end{eqnarray}   
For the total plasmon width in the medium one has to take into account
the longitudinal plasmon creation process with a probability
$\Gamma_{\rm cr}({\cal E})=e^{-{\cal E}/T}\,\Gamma_{\rm ab}({\cal
E})$ so that altogether~\cite{Wel}
\begin{equation}\label{eq:G3}
\Gamma_L=\Gamma_{\rm ab}-\Gamma_{\rm cr} =
\left(1-e^{-{\cal E}/T}\right)\,\Gamma_{\rm ab}.
\end{equation} 
This ``width'' plays the role of the imaginary part of the
polarization operator ${\rm Im}\,\Pi^{L}({\cal E})=-{\cal
E}\,\Gamma_L({\cal E})$.  With these results one finds
for Eq.~(\ref{eq:gamma})
\begin{equation}\label{eq:Gamma2}
\gamma = \frac{2 \alpha}{3}\; \frac{\mu^2}{{\cal E}^2}.
\end{equation}
The temperature dependence has magically canceled!

\subsection{Axion Emissivity}

In order to obtain the plasma's axion emissivity we finally need to
integrate over the initial-state electron phase space as well as the
axion energies,
\begin{equation}\label{eq:Q1}
Q_a=\int\frac{2 d^3 {\bf p}_1}{(2 \pi)^3}\int
d E_a\,\frac{d W}{d E_a}\,E_a\,f (E_1)\,
[1-f (E_2)],
\end{equation}
where $E_2=E_1-E_a$ and $f(E)=(e^{( E - \mu)/T} + 1)^{-1}$ is the
electron's Fermi-Dirac distribution function.
For a degenerate plasma with $\mu\gg T$ the integral over the initial
electron energy $E_1$ can be easily calculated,
\begin{equation}\label{eq:Int}
J=\int dE_1\,f(E_1)\,[1-f(E_2)]\,R(E_1) \simeq
\frac{E_a \; R(\mu)}{e^{E_a/T} - 1}.
\end{equation}
With this result the emissivity is
\begin{equation}\label{eq:Q2}
Q_a\simeq\frac{\mu^2}{2\pi^2}\,\int^{\pi}_0\sin\theta\,d\theta
\int_0^\infty\frac{d E_a\,E_a^2}{e^{E_a/T}-1}\, 
\frac{d W}{d E_a}\,.
\end{equation}
Taking into account Eqs.~(\ref{eq:W1}) and~(\ref{eq:Gamma2}), the
integration over the angle $\theta$ and the axion energy $E_a$ gives
\begin{equation}
Q_a \simeq \frac{\bar g_{a\gamma}^2}{48 \pi^2 \alpha}\,
(e B)^2\,\frac{{\cal E}^3}{ e^{{\cal E}/T} - 1}.
\label{eq:Q3}
\end{equation}
For a numerical evaluation note that the combination $eB$ is
independent of the system of electromagnetic units. In rationalized
natural units we have $\alpha=e^2/4\pi\simeq 1/137$ and
the field strength of $1\,\rm Gauss$ corresponds to 
$1.95\times10^{-2}\,\rm eV^2$. 

Equation~(\ref{eq:Q3}) gives us the emissivity of an ultrarelativistic
degenerate plasma in the resonance region.  When the plasma is
strongly degenerate we may have ${\cal E} \gg T$. In this case the
resonant emissivity is exponentially suppressed, so the nonresonant
contribution
\begin{eqnarray}
Q_a \simeq \frac{2 \zeta(5)}{3 \pi^3}\,\bar g_{a\gamma}^2\,
(e B)^2\,\frac{\mu^2 T^5}{{\cal E}^4}
\label{eq:Q4} 
\end{eqnarray}
will dominate.


\section{Plasmon-Axion Conversion}

We have calculated the cyclotron emission process in the resonance
region. It is known that in this case $e^-\to e^-a$ effectively
reduces to two consecutive processes: cyclotron radiation of a real
longitudinal plasmon $e^- \to e^- \gamma_L$ and a subsequent
plasmon-axion transition $\gamma_L\to a$.  Therefore, one may simply
picture the axion emission as a result of the plasmon-axion conversion
in the presence of the magnetic field.

Starting from the effective Lagrangian Eq.~(\ref{eq:Lag}) the matrix
element for the $\gamma_L (q)\to a(q')$ transition can be written as
\begin{equation}\label{eq:S3}
S\simeq\frac{\bar g_{a \gamma}\,(2\pi)^4\delta^4(q-q')}
{\sqrt{2\omega V\,2E_aV}}\, 
\sqrt{Z_L}\,(\ell\tilde Fq),
\end{equation}
where $\ell_\alpha$ is the four vector of the longitudinal plasmon
polarization defined in Eq.~(\ref{eq:G}), 
$q=(\omega,{\bf q})$ is the initial-state plasmon four momentum, and 
$q'=(E_a,{\bf q}')$ the final-state axion four momentum. 
The transition rate is
\begin{equation}\label{eq:W2}
dW=\frac{\pi\,\bar g_{a\gamma}^2}{2 \omega E_a}\,
\delta^4(q - q')\,Z_L\,(\ell\tilde F q)^2 \, d^3{\bf q}',
\end{equation}
leading to an axion emissivity of 
\begin{equation}\label{eq:Q5}
Q_a=\int \frac{d^3 {\bf q}}{(2\pi)^3}\int dW\,
\frac{E_a}{e^{{\omega}/T}-1}\,.
\end{equation}
Integrating the phase space of the initial plasmon and final axion
this becomes 
\begin{equation}
Q_a \simeq \frac{\bar g_{a\gamma}^2}{48\pi^2\alpha}\,
\frac{(e B)^2\,{\cal E}}{e^{{{\cal E}}/T}-1}\,  
\biggl({\frac{q^2 Z_L}{1-d\omega_L/d k} 
\biggr)_{k={\cal E}}}.
\label{eq:Q6}
\end{equation}
Here $\omega_L(k)$ is the plasmon energy as a function of wave-number
(see Fig.~1).  The expression in brackets is always found to be ${\cal
E}^2$, independently of the plasma conditions (relativistic,
nonrelativistic, degenerate, nondegenerate). Therefore, we
immediately recover Eq.~(\ref{eq:Q3}). 

We conclude that the axion emissivity Eq.~(\ref{eq:Q3}), which we
calculated for an ultrarelativistic degenerate plasma, {\em is
actually valid in the general case}.  We only need to identify the
crossing point ${\cal E}$ of the dispersion relations for the relevant
plasma conditions. For a nonrelativistic plasma we find
\begin{equation}\label{eq:E1}
{\cal E}^2 \simeq 4\pi\alpha\,\frac{n_e}{m_e}, 
\end{equation}
where $n_e$ is the electron density.
For an ultrarelativistic nondegenerate plasma it is 
\begin{equation}\label{eq:E2}
{\cal E}^2 \simeq \frac{4 \pi \alpha }{3}\,T^2\,
\left[\ln\left(\frac{4 T}{m_e}\right)+\frac{1}{2}-\gamma_E+
\frac{\zeta'(2)}{\zeta(2)}\right],
\end{equation}
where Euler's constant $\gamma_E$ is $0.577216$ and
$\zeta'(2)/\zeta(2)=-0.569961$ so that
$1/2-\gamma_E+\zeta'(2)/\zeta(2)=-0.647177$.  For an ultrarelativistic
degenerate plasma ${\cal E}^2$ is given by Eq.~(\ref{eq:Dis2}).


\section{Discussion and Summary}

We have calculated the axion emissivity of a plasma in the presence of
a strong magnetic field due to the resonant transition between
longitudinal plasmons and axions. The field was taken to be strong in
the sense $eB\gg\alpha^3 E^2$ where $E$ is a typical electron energy
so that phase space is opened for the cyclotron process $e^-\to e^-
a$, yet it was also taken to be weak in the sense $E^2\gg eB$ so that
the electron energy is still the largest energy scale in the
problem. For the plasma dispersion relation we could then use the
zero-field expressions. We have computed the emission rate as a
cyclotron process $e^-\to e^-a$ with an intermediate plasmon on
resonance, and also directly as a transition of thermally excited
plasmons into axions. Both calculations yield the same result
Eq.~(\ref{eq:Q3}) in terms of the crossing energy ${\cal E}$ of the
axion with the longitudinal plasmon dispersion relation. With the
appropriate expression for ${\cal E}$ this result applies to all
plasma conditions.

While this process and its evaluation are conceptually quite
intruiging, the actual energy-loss rate is rather small.
If we take the conditions in a supernova core after collapse as an
example, an electron chemical potential of $200~{\rm MeV}$ is a
representative value, leading to ${\cal E}=46~{\rm MeV}$. With a
temperature $T=30~{\rm MeV}$ we thus have 
${\cal E}/T=1.53$. We express the axion-photon coupling as
$g_{a\gamma}=m_{\rm eV}/0.69\times 10^{10}~{\rm GeV}$ for typical 
axion models in terms of the axion mass with $m_{\rm eV}=m_a/1~{\rm
eV}$. The emission rate is then
\begin{equation}
Q_a=2.0\times10^{30}~{\rm erg~cm^{-3}~s^{-1}}~
\frac{m_{\rm eV}^2\,B_{16}^2\,\mu_{200}^3}{e^{{\cal E}/T}-1},
\end{equation}
where $B_{16}=B/10^{16}~{\rm Gauss}$ and $\mu_{200}=\mu/200~{\rm
MeV}$. With ${\cal E}/T=1.53$ the denominator is 3.62 so that
$Q_a=0.6\times10^{30}\ {\rm erg~cm^{-3}~s^{-1}}~m_{\rm
eV}^2\,B_{16}^2$. Dividing by a typical mass density of
$\rho=3\times10^{14}~{\rm g~cm^{-3}}$ this is $Q_a/\rho=2\times10^{15}\
{\rm erg~g^{-1}~s^{-1}}~m_{\rm eV}^2\,B_{16}^2$. This is to be
compared with the neutrino energy loss rate of a supernova core during
its first few seconds after collapse of around $10^{19}~{\rm
erg~g^{-1}~s^{-1}}$. Comparing this number with the zero-field
emission rate by nucleon bremsstrahlung processes already gives a
limit of around $m_a\alt 10^{-2}~{\rm eV}$ so that an unrealistically
large magnetic field would be required to cause a significant
plasmon-conversion rate.

We may turn this finding around and conclude that the axion ``window
of opportunity'' remains open. Even huge magnetic fields in a SN core
do not seem to change the usual picture that axions are emitted
primarily by fluctuating nucleon spins.


\section*{Acknowledgments}

This research was partially supported by INTAS under grant No.~96-0659,
by the Deutsche For\-schungs\-ge\-mein\-schaft under grant No.\
SFB~375
and by the Russian Foundation for Basic Research under grant 
No.~98-02-16694. 
N.M.~and L.V.~acknowledge support by the Max-Planck-Institut
f\"ur Physik (Munich) and by DESY (Hamburg) during a visit when this
work was begun.       


\end{document}